\documentclass[reprint,superscriptaddress,showpacs,nofootinbib,aps,prl]{revtex4-1}
\usepackage{amssymb}
\usepackage{amsmath}
\usepackage{graphicx}
\usepackage{dcolumn}
\usepackage{bm}
\usepackage[utf8]{inputenc}
\usepackage[frenchb]{babel}
\usepackage{braket}

\begin{document}

\title{Hanle model of a spin-orbit coupled Bose-Einstein condensate of excitons\\ in semiconductor quantum wells}
\author{S. V. Andreev}
\email[Electronic adress: ]{Serguey.Andreev@gmail.com}
\affiliation{ITMO University, St. Petersburg 197101, Russia}
\affiliation{University of Bordeaux, LOMA UMR-CNRS 5798, F-33405 Talence Cedex, France}
\author{A. V. Nalitov}
\affiliation{Physics and Astronomy School, University of Southampton, Highfield, Southampton SO171BJ, United Kingdom}
\date{\today}

\begin{abstract}
We present a theoretical model of a driven-dissipative spin-orbit coupled Bose-Einstein condensate of indirect excitons in semiconductor quantum wells (QW's). Our steady-state solution of the problem shares analogies with the Hanle effect in an optical orientation experiment. The role of the spin pump in our case is played by boson stimulated scattering into the linearly-polarized ground state and the depolarization occurs as a result of exchange interaction between electrons and holes. Our theory agrees with the recent experiment [A. A. High et al., Phys. Rev. Lett. \textbf{110}, 246403 (2013)], where spontaneous emergence of spatial coherence and polarization textures have been observed. As a complementary test, we discuss a configuration where an external magnetic field is applied in the structure plane.  
\end{abstract}

\pacs{Valid PACS appear here}
\maketitle

Spin-orbit (SO) coupled Bose-Einstein condensation (BEC) has been under intense investigation in atomic gases during the past few years \cite{Pitaevskii}. A key feature of this new type of BEC is the macroscopic occupation of single-particle states with non-zero momentum \cite{Lin2011}, with dramatic consequences for possible quantum phases and transitions between them. Among theoretically predicted are the stripe phase \cite{Li2012, Ketterle}, the thermodynamically stable meron \cite{Wilson}, and the many-body "cat" state \cite{GalitskiReview}. Practical implementation of the latter could pave the way towards fault tolerant quantum computation at room temperature \cite{Galitski2008}.

In solids the bosonic quasiparticles which display the SO interaction are excitons and polaritons - bound electron-hole pairs in a semiconductor media coupled to light \cite{Ivchenko}. Important distinction of these systems from their atomic counterpart is their driven-dissipative character. In most cases polaritonic condensates are strongly out-of-equilibrium, behaving essentially as lasers. In contrast, excitons can establish a kinetic equilibrium with respect to the interactions needed for formation of a Bose-Einstein condensate \cite{Ivanov}. Promising candidates for creation and manipulation of interacting degenerate Bose gases in semiconductors are the so-called indirect excitons, formed of spatially separated electrons and holes \cite{ButovReview}.

Recently, spontaneous onset of extended spatial coherence and polarization textures have been observed in the photoluminescense (PL) of indirect excitons in coupled semiconductor quantum wells (QW's) \cite{High}. The textures at different locations in the QW plane appear simultaneously at the same critical temperature and their orientation is pinned to the crystallographic axes. Whereas the former observation suggests a second-order phase transition \cite{ScaleInvariance}, the latter fact allows one to unambiguously identify the SO interaction as playing a crucial role in these phenomena \cite{High, SpinTextures}. Until now, however, no link between formation of an exciton condensate in the momentum space and the spin texture has been established. 

The present paper is aimed at filling this gap in the understanding of exciton BEC. We propose an ideal gas model of SO coupled exciton condensates accounting for the radiative decay and incoherent pumping from an external reservoir. The emergent physics is reminiscent of the Hanle effect, which consists in suppression of the spin polarization induced by a continuous wave (cw) circularly polarized laser in a transverse magnetic field \cite{Hanle, OO}. In our case, stimulated scattering of excitons into the linearly-polarized ground state (GS) is followed by coherent precession ("depolarization") of the condensate pseudo-spin in an effective magnetic field due to the electron-hole exchange interaction. The stationary configuration (pseudo-spin rotation angle) is determined by the balance between the radiative decay rate and the pumping from the reservoir. The model can be straightforwardly generalized to account for an external magnetic field. In the Faraday configuration (the magnetic field is applied perpendicularly to the structure plane) our theory reproduces the rotation of the linear polarization texture observed in the experiment \cite{High}. As a complementary test of our model, we discuss the Voigt geometry (in-plane magnetic field).

Our starting point is the following single-particle SO-coupled Hamiltonian
\begin{equation}
\label{SPHamiltonian}
\hat H=\left( \begin{array}{cccc}
E_k & \beta_{h} ke^{-i\varphi} & \beta_{e} ke^{i\varphi} & 0 \\
\beta_{h} ke^{i\varphi} & E_k & 0 & \beta_{e} ke^{i\varphi}\\
\beta_{e} ke^{-i\varphi} & 0 &E_k & \beta_{h} ke^{-i\varphi}\\
0 & \beta_{e}ke^{-i\varphi} & \beta_{h}ke^{i\varphi} & E_k 
\end{array} \right),
\end{equation}
which is obtained by the Kronecker summation of the corresponding Dresselhaus Hamiltonians for the constituent electrons and holes \cite{Andreev2}. Here $E_k=\hbar^2 k^2/2m$ is the exciton kinetic energy and $\varphi$ is the angle between the exciton wavevector $\bm k$ and the $[100]$ crystal axis (chosen as the $x$-axis). The lowest energy eigenstates of the Hamiltonian \eqref{SPHamiltonian} are the plane waves, characterized with the dispersion
\begin{equation}
\label{dispersion}
E(k) = \frac{\hbar^2 k^2}{2m} - k \left( \beta_e + \beta_h \right)
\end{equation}
and the following spin structure
\begin{equation}
\label{u}
\ket{\mathrm{GS}}={1\over 2}\left( \begin{matrix}
1, &-e^{ \mathrm{i} \varphi}, & -e^{-\mathrm{i} \varphi}, & 1
\end{matrix} \right)^\mathrm{T}.
\end{equation}
The dispersion \eqref{dispersion} has a characteristic minimum at  $k_0=m ( \beta_e + \beta_h )/\hbar^2$. The  $\varphi$-dependent part $\bra{\varphi}=1/2(-e^{-i\varphi},-e^{i\varphi})$ in $\ket{\mathrm{GS}}$ corresponds to the bright exciton state and can be used to construct a pseudo-spin $\bm S=1/2\bra{\varphi}\bm\sigma\ket{\varphi}$ which defines the polarization of the light emitted by the exciton (here $\bm\sigma$ is the Pauli vector). Thus, the linear polarization degrees are given by $\rho_{l}\equiv (I_{x}-I_{y})/(I_{x}+I_{y})=2 S_x/\braket{\varphi|\varphi}$
and $\rho_{l'}\equiv (I_{x'}-I_{y'})(I_{x'}+I_{y'})=2 S_y/\braket{\varphi|\varphi}$ with the frame $(x',y')$ being rotated by $\pi/4$ with respect to $(x,y)$. One can easily see that the light emitted by the state \eqref{u} is polarized along the wavevector $\bm k$.

In the work \cite{High} a vortex of linear polarization and a four-leaf circular polarization texture have been observed around the localized bright spots (LBS's) in the exciton PL pattern. The bright spots are characterized by high intensities and partially suppressed coherence of the exciton PL. As has been recently pointed out by one of us \cite{Andreev2}, an LBS surrounded by a coherent spin-polarized halo represents an exciton cloud trapped in an external potential. The potential can be assumed to be of a harmonic type
\begin{equation}
\label{trap}
V_{\mathrm{ho}}(x,y)=\frac{m(\omega_x^2x^2+\omega_y^2y^2)}{2},
\end{equation}
with $\omega_{x,y}$ being the harmonic oscillator frequencies. Below we present an ideal gas model of an SO-coupled exciton condensate in the 2D harmonic trap \eqref{trap}, where we let $\omega_x=\omega_y\equiv\omega$ for simplicity. The model reproduces the main features observed in the experiment. 

The two-body interactions, while not being explicitly taken into account in our calculations, ensure the important relation
\begin{equation}
\label{condition1}
\tau_{\bm{k}}\ll\tau
\end{equation}
between the characteristic momentum relaxation time $\tau_{\bm{k}}$ and the lifetime $\tau$ of excitons. In an SO-coupled gas of semiconductor electrons the elastic collisions are known to play an important role in the relaxation of spin according to the Dyakonov-Perel scenario \cite{Dyakonov, Glazov}. Frequent collisions slow down the relaxation of an artificially created spin polarization. In our model, instead of an external optical pumping we assume a cold reservoir of unpolarized electrons and holes, which bind into (initially) unpolarized excitons. In a bosonic ensemble of excitons being in dynamical equilibrium with the cold bath the condition \eqref{condition1} favours spontaneous formation of a metastable Bose-Einstein condensed phase \cite{Pitaevskii}. Crucially, due to the coupling of the exciton spin to the momentum, encoded in the Hamiltonian \eqref{SPHamiltonian}, thermalization and condensation occurs both in momentum and spin subspaces. As a result, an equilibrium GS has more complex structure than just the usual lowest-energy state of the harmonic oscillator, typical for the potential \eqref{trap}. In the limit of strong SO coupling $k_0 l_{\mathrm{ho}}\gg 1$, where $l_{\mathrm{ho}}=\sqrt{\hbar/m\omega}$ is the harmonic oscillator length, the GS wavefunction can be obtained as follows. In the momentum space representation the harmonic potential $V_{\mathrm{ho}}(x,y)$ can be regarded as a kinetic energy operator of a fictitious particle moving in the potential energy landscape \eqref{dispersion}. The corresponding Schrodinger equation reads
\begin{equation}
\left[-\frac{m\omega^2\nabla_{\bm k}^2}{2}+E(k)\right]\Psi(\bm k)=E \Psi(\bm k).
\end{equation}
By using the substitution $\Psi_l(\bm k)=e^{\mathrm i l\varphi}f_l(k)\ket{\mathrm{GS}}/\sqrt{k}$ with $\ket{\mathrm{GS}}$ given by \eqref{u} and integrating out the angular degree of freedom, we obtain
\begin{equation} 
\label{eq_radial}
{m \omega^2 \over 2} \left[ - {\mathrm{d}^2 \over \mathrm{d} k^2 } + {1 \over k^2} \left( l^2 + {1 \over 4} \right) \right] f_l = \left( E_l - E(k) \right)f_l.
\end{equation}
Developing the radial motion in equation \eqref{eq_radial} in the vicinity of $k_0$ we have for the eigenenergies:
\begin{equation}
E_{n,l} = E(k_0) + {m \omega^2 \over 2 k_0^2} \left(l^2 + {1 \over 4} \right) + \hbar \omega \left( {1 \over 2} +  n \right),
\end{equation}
where $n$ and $l$ are the radial and the orbital integer quantum numbers, respectively. The new ground state corresponds to $n=l=0$ and is described by the following wavefunction
\begin{equation}
\label{PsiK}
\Psi_0(\mathbf{k})=\pi^{-1/4}\sqrt{\frac{l_{\mathrm{ho}}}{2\pi k}}\exp \left( - { \hbar (k - k_0)^2 \over 2 m \omega} \right) \ket{\mathrm{GS}}.
\end{equation}
In the coordinate space the GS wavefunction reads
\begin{equation}
\label{PsiR}
\Psi_{0}(\mathbf{r})=\frac{1}{\sqrt{2\pi k_0}}\left( \begin{array}{c}
J_0 (k_0 r)\\
J_1 (k_0 r) e^{\mathrm{i}(\phi-\pi/2)}\\
J_1 (k_0 r) e^{-\mathrm{i}(\phi-\pi/2)}\\
J_0 (k_0 r) 
\end{array} \right),
\end{equation}
with $\phi$ now being the polar angle of the radius-vector $\mathbf{r}$. The spin structure of the ground state \eqref{PsiK} at the angle $\varphi$ in the $\bm k$-space is reproduced in \eqref{PsiR} at the angle $\phi=\varphi+\pi/2$. The pseudospin $\bm S$ constructed from the bright part of \eqref{PsiR} lies in the QW plane and rotates by $4\pi$ when going around the LBS. In terms of the quantities $\rho_{l}$ and $\rho_{l'}$ this corresponds to the $2\pi$ vortex of linear polarization observed in the experiment [Fig. \ref{Fig1} (a)].

In order to obtain a finite circular polarization $\rho_c=2S_z$ we need to go beyond the equilibrium model \eqref{SPHamiltonian} and account for the driven-dissipative nature of the system. In fact, our assumption of the kinetic equilibrium \eqref{condition1} by no means implies a full thermodynamic equilibrium. The SO coupling results in formation of a linearly-polarized condensate \eqref{PsiK}, but relaxation with respect to $\bm k$-independent spin interactions may take time longer than the exciton lifetime $\tau$. Among those latter type of interactions are the long-range electron-hole exchange and the Zeeman interaction with an external magnetic field. These interactions are thus expected to yield a dynamical correction
\begin{widetext}
\begin{figure*}[t]
\centering
\includegraphics[width=1.6\columnwidth]{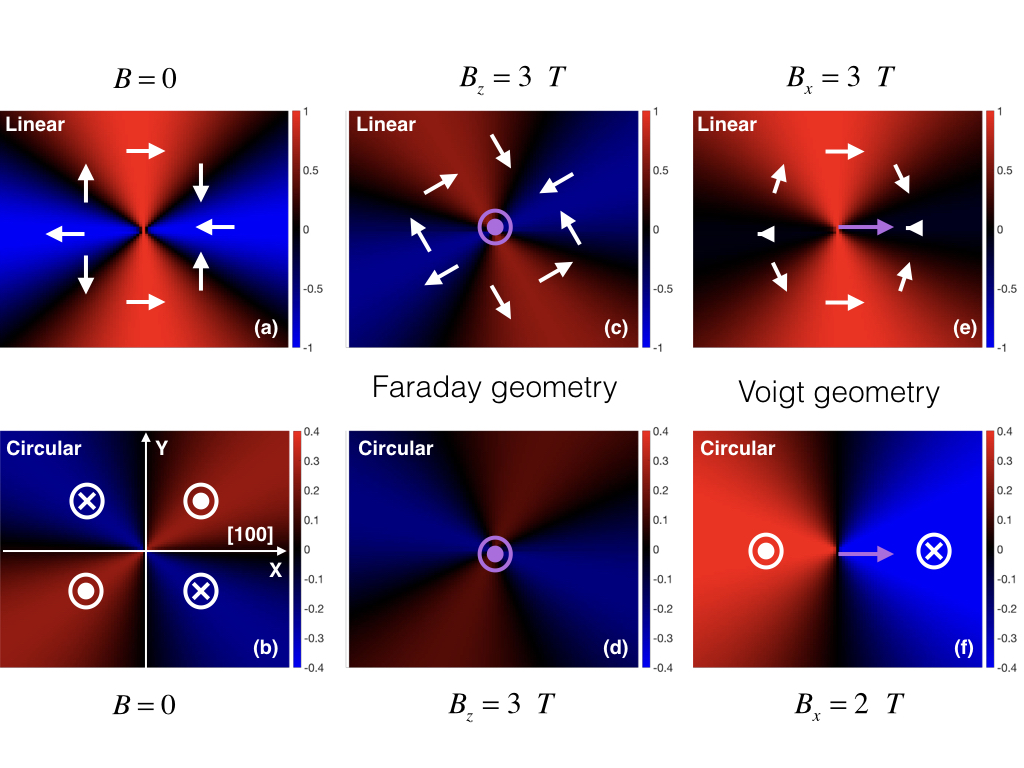}
\caption{Textures of the linear [$\rho_l$, shown in (a,c,e)] and circular [$\rho_c$, shown in (b,d,f)] polarization degrees of the light emitted by an exciton condensate in a trap. The coordinate axes [shown by white arrows in (b)] are chosen along the main crystallographic axes of the sample, consistently with the experiment \cite{High}. For $B=0$ the condensate pseudo-spin $\bm S$ [white arrows in (a)] rotates by $4\pi$ when going around the trap center. This corresponds to the $2\pi$ vortex of linear polarization observed in the experiment. The electron-hole exchange acts as an effective magnetic field acting on $\bm S$ along the $x$-axis. The $y$-component of $\bm S$ precesses around this field. In the steady-state this results in appearance of domains with positive and negative circular polarization [the four-leaf polarization texture in (b)], which corresponds to $S_z>0$ (schematically depicted by $\odot$) and $S_z<0$ (depicted by $\otimes$). The magnetic field applied along $z$ (purple $\odot$, $B_z>0$) rotates the linear polarization texture (c) and suppresses the circular polarization $\rho_c$ (d). In contrast, an in-plane magnetic field applied along $x$ (purple arrow) enhances $|\rho_c|$ [simultaneously transforming the four-leaf pattern into a two-leaf one, (f)] and destroys the negative part of $\rho_l$ (e). We take $\tau_c=0.1$ ns, $\beta_e=\beta_h=0.5$ $\mu eV\times\mu m$ (note, that these are the values of the Dresselhaus constants renormalized by the corresponding mass ratios \cite{Andreev2}), $g_e=0.1$, $g_h=0.15$, $\delta=1$ $\mu eV$ and $S_0=1$.}
\label{Fig1}
\end{figure*} 
\begin{equation}
\label{correction}
\hat H'=\left( \begin{array}{cccc}
-\frac{1}{2}g_{2}\mu_B B_z & 0 & -\frac{1}{2}g_e\mu_B B_x & 0 \\
0 & \frac{1}{2}g_{1}\mu_B B_z & -\delta &  -\frac{1}{2}g_e\mu_B B_x\\
 -\frac{1}{2}g_e\mu_B B_x & -\delta & -\frac{1}{2}g_{1}\mu_B B_z & 0\\
0 &  -\frac{1}{2}g_e\mu_B B_x & 0 & \frac{1}{2}g_{2}\mu_B B_z 
\end{array} \right)
\end{equation}
\end{widetext}
to the Hamiltonian \eqref{SPHamiltonian}. Here $g_{1}=g_h-g_e$ and $g_2=g_e+g_h$ with $g_{e,h}$ being the electron (hole) g-factors, $\mu_B$ is the Bohr magneton and $\delta$ is the long-range exchange constant \cite{Bir}. The in-plane heavy-hole g-factor is assumed to be equal to zero (we neglect possible random stresses typical for the samples of moderate quality \cite{Andreev4}).

A proper description of the spin dynamics of a 4-component system in the presence of decay and coherent pumping from a reservoir can be done by using the Lindblad equation for the spin density matrix
\begin{equation}
\label{Liouville}
i\hbar\frac{d\hat\rho}{dt}=[\hat H+\hat H',\hat\rho]+\frac{i\hbar S_0}{\tau_c}\hat\rho_{\mathrm{GS}}-\frac{i\hbar}{\tau_c}\hat\rho,
\end{equation}
where $\tau_c$ is the condensate lifetime (in general, $\tau_c<\tau$),
\begin{equation}
\hat\rho_{\mathrm{GS}}=\ket{\mathrm{GS}}\bra{\mathrm{GS}}
\end{equation}
with $\ket{\mathrm{GS}}$ being defined by \eqref{u}, and $S_0/\tau_c$ is the stimulated scattering rate being proportional to the number of particles in the condensate. In Eq. \eqref{Liouville} we assume the condition \eqref{condition1} and
\begin{equation}
\tau\ll\tau_s,
\end{equation}
where $\tau_s$ is the exciton spin coherence time. In the steady-state configuration one has
\begin{equation}
\label{steadystate}
\frac{d\hat\rho}{dt}=0,
\end{equation}
which yields a system of 16 coupled linear equations on the matrix elements of $\hat\rho$. The 3 components of the exciton pseudo-spin and the related linear and circular polarization degrees can then be obtained by using the formulas 
\begin{equation}
\begin{split}
2S_x&=I\rho_{l}=\rho_{-1+1}+\rho_{+1-1},\\
2S_y&=I\rho_{l'}=\mathrm{i}(\rho_{-1+1}-\rho_{+1-1}),\\
2S_z&=I\rho_{c}=\rho_{+1+1}-\rho_{-1-1},
\end{split}
\end{equation}
where $I=\rho_{-1-1}+\rho_{+1+1}$. 

The results of solution of Eq. \eqref{steadystate} are presented in Fig. \ref{Fig1}. The observed phenomenology is reminiscent of the Hanle effect in an optical orientation experiment \cite{OO}. The role of the spin pump in our case is played by the boson stimulated scattering into the linearly-polarized GS [Fig. \ref{Fig1} (a)]. The long-range exchange interaction between the electrons and holes acts as an effective magnetic field of the magnitude $2\delta$ oriented along the $x$-axis in the structure plane. This field rotates the condensate pseudo-spin, so that the latter acquires a non-zero component $S_z$ along the growth direction [Fig. \ref{Fig1} (b)]. The rotation angle in the steady-state configuration \eqref{steadystate} is determined by the balance between the radiative decay of the condensate and the pump from the reservoir. An external magnetic field applied along the $z$-axis (Faraday geometry) rotates the pseudo-spin component lying in the structure plane [Fig. \ref{Fig1} (c)], which yields the rotation of the linear polarization vortex observed in the experiment \cite{High}. 

The effect of an in-plane magnetic field (Voigt geometry) is more subtle. Measurements of the exciton polarization in this configuration could provide a good test of our theory. As an example we show transformation of the linear [Fig. \ref{Fig1} (e)] and circular [Fig. \ref{Fig1} (f)] polarization textures by a magnetic field along the $x$-axis with $B_x=3$ T and $B_x=2$ T, respectively. The field enhances the absolute degree of the circular polarization $|\rho_c|$ [simultaneously transforming the four-leaf spatial pattern into a two-leaf one] and destroys the negative part of $\rho_l$.

Let us now briefly discuss a possible effect of the two-body interactions on the properties of the ground state. A net repulsive interaction is expected to result in population of the states with larger angular momentum $l$ and extrusion of the condensate to the peripheric region of the trap as to reduce the density in the center \cite{Santos}. On the other hand, if one of the interaction channels admits a resonance, an increase of the chemical potential in the center may result in formation of a condensate of biexcitons, distinguished from the exciton condensate by a suppressed coherence of the emitted light \cite{Andreev3}. In practice, both effects may be present simultaneously, complementing each other. An interplay between the SO interaction and resonant pairing of bosons is a challenging question which will be addressed in our future studies of many-body quantum phases of excitons.

To conclude, we have shown that the experimentally observed polarization textures in the PL of indirect excitons can be interpreted in terms of the SO-coupled exciton Bose-Einstein condensation. Our theory suggests a new mechanism of spontaneous formation of the spin polarization in semicondutors, which one can dub "\textit{boson stimulated spin pumping}". In the absence of full thermodynamic equilibrium the dynamics of the condensate spin can be described by the Hanle-like equation on the spin density matrix \eqref{Liouville}. The model can be adopted to the atomically thin layered materials \cite{Novoselov}, as well as to high-quality semiconductor microcavities \cite{Snoke}.     

S.V. thanks M. Glazov for his remarks. This work has been supported by the Government of the Russian Federation (Grant 074-U01) through ITMO Postdoctoral Fellowship scheme.
       
\preprint{APS/123-QED}

\end{document}